\documentclass[journal,10pt]{IEEEtran}
\usepackage{graphicx, epstopdf}
\usepackage{amssymb, amsmath}
\usepackage{commath}
\usepackage{setspace}
\usepackage{acronym}
\usepackage{mathrsfs}
\usepackage{ifthen}
\usepackage{textcomp}
\usepackage{float}
\usepackage{subfigure}
\usepackage[table]{xcolor}
\usepackage{color,soul}
\usepackage{cite}
\usepackage{setspace}
\usepackage{mathtools}
\usepackage{bm}
\usepackage{url}
\usepackage{algorithm}
\usepackage{algpseudocode}
\usepackage[switch]{lineno}

\IEEEoverridecommandlockouts
\begin{document}

\title{The Spatial Ecology of War and Peace}

%\doublespacing
%\linenumbers

\author{Weisi Guo\textsuperscript{1,2*}, Xueke Lu\textsuperscript{2}, Guillem Mosquera Do\~nate\textsuperscript{3}, Samuel Johnson\textsuperscript{3}

\thanks{\textsuperscript{1}The Alan Turing Institute, United Kingdom. \textsuperscript{2}School of Engineering, University of Warwick, United Kingdom. \textsuperscript{3}Warwick Mathematics Institute and Centre for Complexity Science, University of Warwick, United Kingdom.}}

\markboth{current preprint version (v3) June 2017 (v1 Apr 2016)}
{Submitted paper}
\maketitle

%\subtitle{\textbf{Summary:} A global network of city interactions can accurately predict the location of conflict zones in modern history, highlighting the importance of geography in war.\\}

\begin{abstract}
Human flourishing is often severely limited by persistent violence. Quantitative conflict research has found common temporal \cite{Zhang07, Hsiang11} and other statistical patterns in warfare \cite{Bohorquez09}, but very little is understood about its general spatial patterns. While the importance of topology in geostrategy has long been recognized \cite{Kaplan12, Blouet05}, the role of spatial patterns in determining a region's vulnerability to conflict is not well understood. Here, we use network science to show that global patterns in war and peace are closely related to the relative position of cities in a global interaction network. We find that regions with \textit{betweenness centrality} above a certain threshold are often engulfed in entrenched conflict, while a high \textit{degree} correlates with peace. In fact, betweenness accounts for over 80\% of the variance in number of attacks. This metric is also a good predictor of the distance to a conflict zone and can estimate the risk of conflict. We conjecture that a high betweenness identifies areas with {fuzzy cultural boundaries} \cite{Lim07}, whereas high degree cities are in cores where peace is more easily maintained. This is supported by a simple agent-based model in which cities influence their neighbours, which exhibits the same threshold behaviour with betweenness as seen in conflict data. These findings not only shed new light on the causes of violence, but could be used to estimate the risk associated with actions such as the merging of cities, construction of transportation infrastructure, or interventions in trade or migration patterns.
\end{abstract}

\section{Introduction}

Conflict has strong spatial components that are not fully understood and cannot be explained by current modeling techniques. Whilst war-torn countries face common socioeconomic and geopolitical challenges, different regions experiences different orders of magnitude in violence. This paper presents an elegant network science framework for understanding the role of global spatial interactions on local violence intensity. Throughout history, most violent incidents have been between culturally or ethnically distinct groups \cite{Denton68, Horowitz00}. Whilst cultural diversity does not in itself cause violence, it can exasperate existing vulnerabilities, especially when there are neighbouring forces with competing interests \cite{Esteban12}. In particular, conflicts have been observed to occur at fuzzy cultural boundaries that nest between coherent cultural groups \cite{Lim07, Legewie16, Tullberg97, Kaufmann98}. Whilst data can be collected on ethnolinguistic and cultural fractionalization \cite{Alesina03}, understanding the interaction forces that support fuzzy cultural boundaries can provide mechanisms to mitigate violence in the long term. One way of collecting data on cultural communities is through social interaction networks \cite{Blondel08}, and interaction networks between communities have been used to model the projection of cultural influence/threat \cite{Baudains16}. However, although it is known that physical geography underpins community formation and influence/threat projection in conflicts \cite{Blondel08, Kaplan12}, there is no universal model able to quantify the role of spatial patterns in war and peace. 
\begin{figure}[t]
    \centering
    \includegraphics[width=1.00\linewidth]{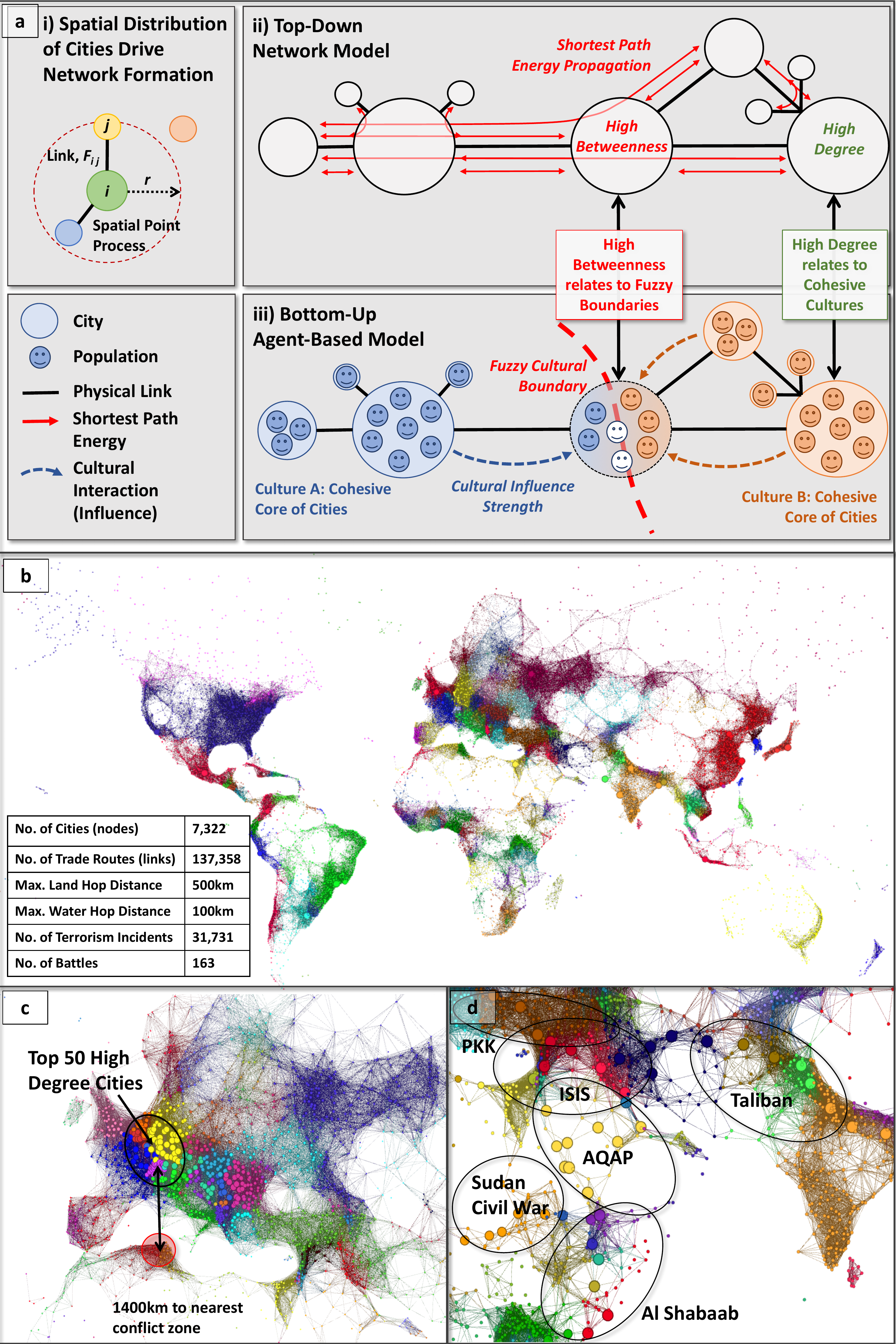}
    \caption{{\bf Complex network of cities and cultural interaction links.} \textbf{a-i)} Cities' spatial location is used to form an interaction network via the gravity law and a hard-disk constraint - the links enable the flow of cultural influence, \textbf{a-ii)} Top-down network model can detect high betweenness cities using shortest-path energy propagation, and \textbf{a-iii)} Bottom-up agent simulations can reveal fuzzy culture boundary cities and cohesive culture cores. The high betweenness metric relates to fuzzy culture and high degree relates to cohesive cultured cities. \textbf{b)} Global interaction network of cities $v \in V$ (colour indicates country and size indicates population). \textbf{c)} Example of high degree cities (size $\propto D(v)$) in Europe, which are far from any major conflicts. \textbf{d)} Examples of high betweenness cities (size $\propto B(v)$) in the Middle East, which are close to major conflicts and international terrorist groups.}
    \label{fig:1}
\end{figure}

In this study, we consider both non-state terrorism and conventional warfare from 2002 to 2014, for which there is spatially accurate reporting of violent events. We use a simple connectivity law to connect neighbouring cities as a proxy for multiplexed interactions. We go on to show that the spatial network of cities is closely related to violence, and propose a simple model to account for the phenomenon.
\begin{figure*}[t]
    \centering
    \includegraphics[width=1.00\linewidth]{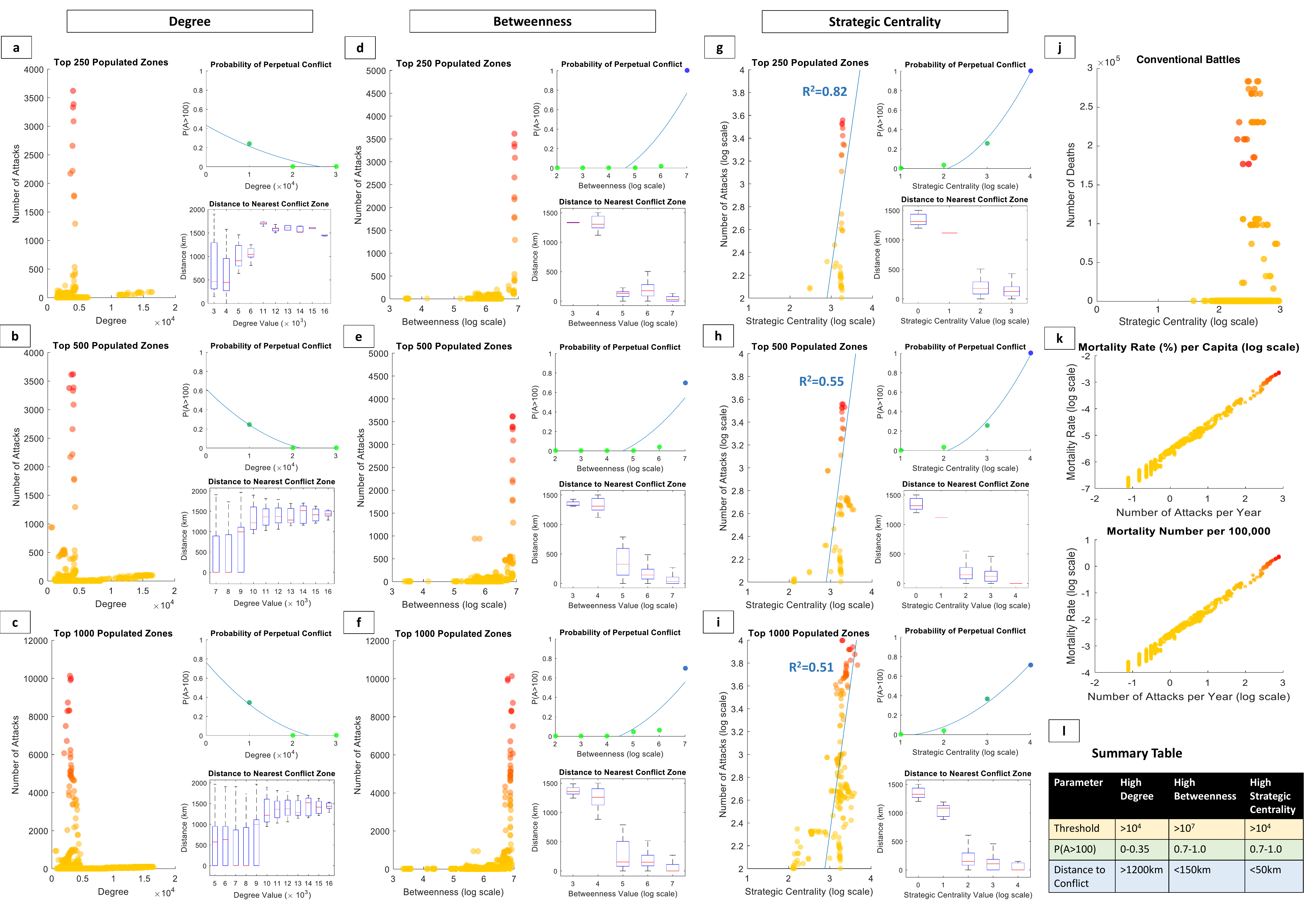}
    \caption{{\bf Scatter plot of number of attacks versus the network metrics of the top populated zones $z$; with smaller panel plots on the probability of suffering a major attack and the distance distribution to the nearest major conflict site.} Subplots are divided in accordance to the top 250, 500, and 1000 populated zones (representing 15\%, 27\%, and 45\% of the total modelled population). The colour gradient of the scatter plots indicate the mortality rate, with darker spots indicating a disproportionately higher mortality rate. Subplots a)-c) show the threshold relationship for the \textit{degree} of a zone and the number of attacks -- high degree ($D(z)>10^4$) experiences almost no attacks and is $>1200$ km away from the nearest major conflict. Subplots d)-f) show the threshold relationship for the \textit{betweenness} of a zone -- high betweenness ($B(z)>10^{7}$) has a high probability of major conflict ($P(A(z)>100)>$0.7), and is close to existing conflict zones ($<$150 km). Combining \textit{degree} and \textit{betweenness}, we propose \textit{strategic centrality} $S(z) = B(z)/D(z)$. Subplots g)-i) show the relationship between the strategic centrality and the number of attacks. The results indicate a strong correlation (R$^{2}<$0.82). A zone with high strategic centrality has a high probability of being in major conflict ($P(A(z)>100)>$0.7) and is close to existing conflict zones ($<$50 km). Subplot j) shows conventional battle death-toll vs. strategic centrality, demonstrating a similar threshold behaviour. Mortality rates can be found in subplot k) and a summary of the findings can be found in subplot l).}
    \label{fig:2}
\end{figure*}

\section{Results}

Using the spatial data from city locations worldwide, we infer a global interaction network via the gravity law, as shown in \textbf{Fig.~\ref{fig:1}a-i} (see \textbf{Methods-B}). This top-down network abstraction uses shortest-path energy propagation (\textbf{Fig.~\ref{fig:1}a-ii} -- \textbf{see Methods-C}) to mirror the complex culture interactions simulated by the bottom-up agent-based model (\textbf{Fig.~\ref{fig:1}a-iii} -- \textbf{see Methods-D}). We use two standard network measures: degree $D$ (number of neighbouring connected cities) and betweenness $B$ (number of shortest paths which pass through the city in question).

We find a robust empirical relationship between conflict and network measures: high degree cities are peaceful and far from the nearest conflict zones, whereas high betweenness cities are often engulfed in persistent violence. Strikingly, there is a threshold effect such that only cities above a certain betweenness are at risk. These results seem consistent with the interpretation that dense cores of high degree nodes in the network correspond to culturally cohesive regions, while high betweenness nodes (which usually link such cores) signify fuzzy cultural boundaries associated with violence. Examples of the highest degree cities are in Western Europe and are far from the nearest conflict zones (\textbf{Fig.~\ref{fig:1}c}). Examples of the highest betweenness cities are in the Middle East, and have experienced high levels of terrorism and conventional violence (\textbf{Fig.~\ref{fig:1}d}). A simple agent-based model of cities which can adopt different states (cultures) and exert an influence on the states of their neighbours reinforces this view: a similar pattern arises, such that nodes in the cores maintain their states while those above a threshold of betweenness tend to flip at high frequency from state to state.

\subsection{Top-Down Network Results}

Conflicts often occur across regions which include several cities. We therefore group cities into small zones $z$, each covering 0.5\% of the global land surface area (500km radius). \textbf{Fig.~\ref{fig:2}} shows scatter plots of the number of terrorist attacks $A(z)$ against the network metrics, with panel plots for both the probability of being a major conflict zone, and the distance to the nearest major conflict zone. A major conflict zone is one that experiences persistent conflict, which is defined as having suffered over 100 attacks between 2002 and 2014 (equivalent to at least 78\% of the years being under attack -- see Supplementary Information (SI)). Isolated high profile terrorist attacks (e.g. 9/11 in New York and 7/7 in London) do not necessarily indicate major conflict zones. This is because the death-toll variance for individual attacks depends on certain on-the-day factors, which do not reflect the general level of threat faced by a region (\textbf{Fig.~\ref{fig:2}k}).

\subsubsection{Degree and Betweenness Centrality}
The results for degree and betweenness centrality indicate a threshold behaviour, whereby if the zone has high degree ($D(z)>10^4$ links -- see \textbf{Fig.~\ref{fig:2}a-c}) or low betweenness ($B(z)<10^7$ shortest paths -- see \textbf{Fig.~\ref{fig:2}d-f}), then there are very few attacks ($<$1/year). Conversely, if the zone has low degree ($D(z)<10^4$) or high betweenness ($B(z)>10^7$), then there is a high probability that the zones will experience major conflict (see \textbf{Fig.~\ref{fig:2}a-f}). We also show that the average distance from any zone to the nearest top-100 major conflict zone rises with increasing degree. High degree zones are at least 1200 km away from the nearest conflict zone, whereas high betweenness zones are usually less than 150 km away. 

\subsubsection{Strategic Centrality}
In order to further refine the statistical prediction of conflict, we define the \textit{strategic centrality} of a zone $z$ as $S(z) = \frac{B(z)}{D(z)}$, which normalizes the betweenness of a city by its number of neighbours. It captures the path importance of a city, since a low degree means fewer alternative paths. The results indicate that zones with a high strategic centrality suffer both a high number of attacks and a dispassionately high mortality rate (see \textbf{Fig.~\ref{fig:2}g-i}). The relationship again displays a threshold effect, which can nevertheless be approximated quite well with a power law for the purpose of prediction. The best-fit power law for the number of attacks $A^{*}$ with respect to the strategic centrality of zones is given by $\log_{10}(A^{*}(z)) \simeq a\log_{10}{S(z)} + b$, where the parameters are $a=4$ and $b=-9$. The corresponding adjusted R-squared value is 0.82 for the top 250 populated zones. Different logistic regression algorithms can increase the R-squared value to 0.92, but it is worth noting that the linear regression is only to serve as a demonstration of predictive powers and not as an essential element of the network model itself. In general, the results show that strategic centrality is a better predictor of conflict than either degree or betweenness. A low strategic centrality zone ($S(z)<10^4$) will experience almost no attacks. On the other hand, high strategic centrality zones ($S(z)>10^4$) are on average less than 50 km away from the nearest major conflict zone. A summary of the findings can be found in the table in \textbf{Fig.~\ref{fig:2}l}. To demonstrate the wider applicability of the approach to conventional conflicts (approx. 150 over the time period compared to 30,000 terrorist and insurgency attacks), \textbf{Fig.~\ref{fig:2}j} shows conventional battle death-toll vs. strategic centrality, demonstrating a similar threshold behaviour. 

To confirm that these results are not spurious, we perform two analyses (see SI). First, we show that similar statistical results to \textbf{Fig.~\ref{fig:2}} for different zone sizes and flow weights, demonstrating model robustness. Second, we show that the network centrality metrics presented here are not proxies of key geopolitical or socioeconomic metrics. The results indicate that strategic centrality is a far superior predictor of conflict (adjusted $R^{2}=0.51$) than any established geo-political or socioeconomic metric considered here (adjusted $R^{2}=0.00-0.13$) and strategic centrality itself is not related to any of these indicators (adjusted $R^{2}=0.01-0.30$). The results show that zones and their cities that are simply near the equator or between densely populated sub-continents do not always have a high betweenness or strategic centrality. Simplified country or county/state connection maps are therefore not as informative as city level network descriptions. The metrics developed in this paper cannot be obtained without considering the city network, and do not appear to be direct proxies for established socioeconomic or geo-political metrics.

\subsubsection{Effect of Cities Merging or Fragmenting}

We further expand the analysis by considering how different aspects of city development would affect strategic centrality. We consider a high betweenness city that connects $M$ cohesive city cores, each with $N$ cities (see Methods-C and Fig.~\ref{fig:M1}). Its strategic centrality is therefore proportional to $(M-1)N$. If the high betweenness city fragments from 1 to $K$ smaller cities, its strategic centrality $S(v)$ decreases in proportion to $(M-1)N/2K$, suggesting that the emergence of new cities around existing high betweenness cities, with connecting links to the cohesive cores, would effectively reduce vulnerability to conflict. 
\begin{figure}[t]
    \centering
    \includegraphics[width=1\linewidth]{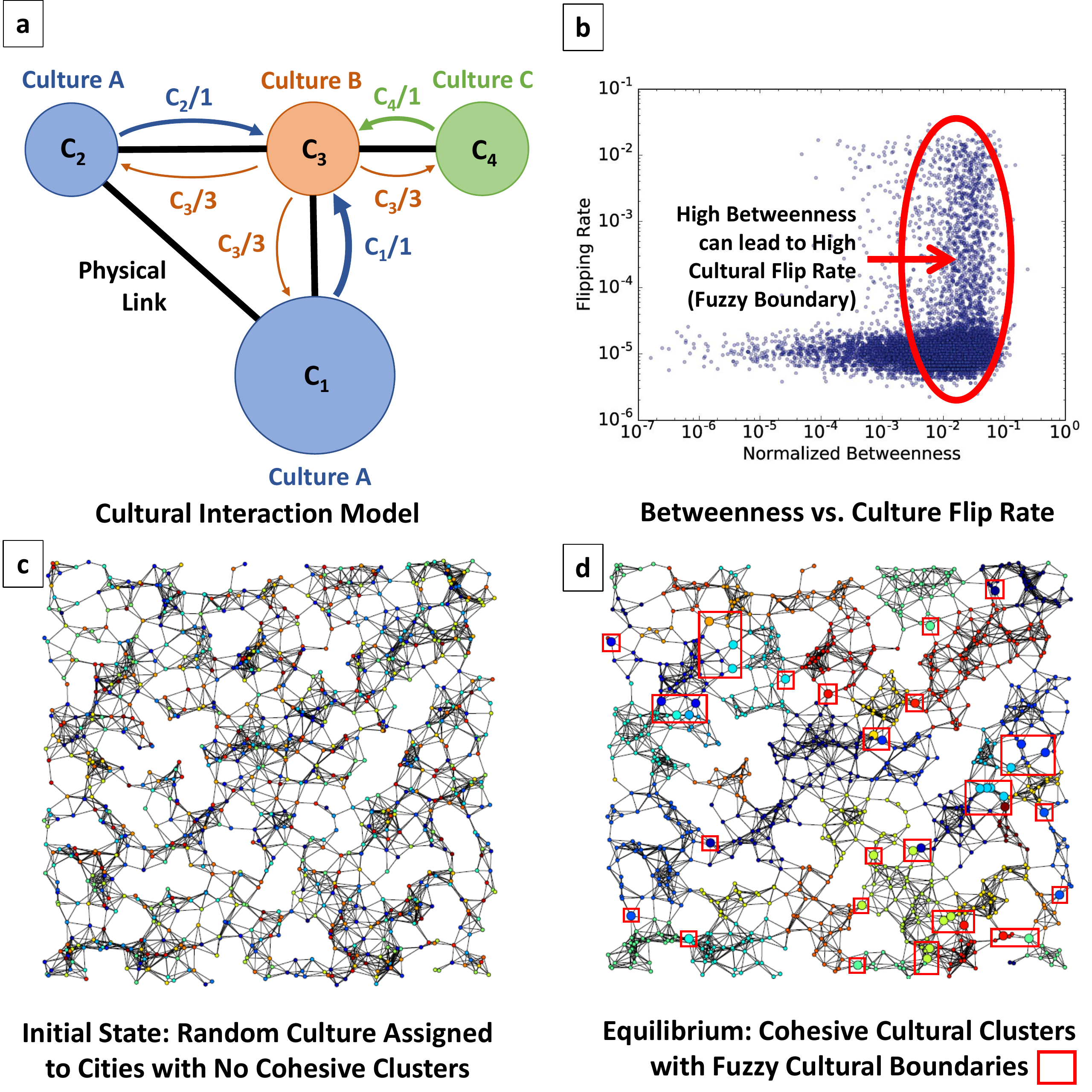}
    \caption{{\bf Agent-Based-Model (ABM) of cities that culturally influence each other through the spatial network.} In subplot a), we illustrate how two cities with culture A, and one city with culture C, exert a strong influence on an in-between city with culture B; whilst that city exert weaker influences in return. In subplot b), we show Monte Carlo simulation results for the agent-based model (ABM), where average flip rate is plotted against betweenness; we can observe that many cities above a threshold of betweenness have a high flip rate. In subplot c), we show that the initial conditions assign each city a random culture; and, in subplot d), cohesive culture cores are formed with fuzzy cultural boundaries (high culture flip rate).}
    \label{fig:3}
\end{figure}

\subsection{Bottom-Up Agent-Based Simulation Results}

The empirical results described above suggest that cities exert an influence on their neighbours, and that geography can determine to a large extent whether a city will converge in some way with its neighbours, or find itself torn between competing influences. In order to study this mechanism, we propose a simple bottom-up agent-based model (ABM). This is an extension of a model previously used to show how governments can influence each other with a view to achieving global cooperation on issues such as climate change \cite{Johnson15}. We assume that each city is characterized at each moment by a 'state', which could represent the reigning government or the dominant culture. Cities are connected by a spatial network, as in the empirical case above, and each city has the capacity to influence the states of its neighbours. This influence, which might represent cultural diffusion or military threat, is such that a city will tend to make its neighbours adopt the same state as itself.

Each city's capacity $C_{i}$ to project influence scales linearly with its population and inversely with the number of neighbours currently in different states to itself (see \textbf{Fig.~\ref{fig:3}a}). At each time step of the simulation, cities update their states according to the sum influence towards each possible state. This model is similar to others used to study social interaction, such as Axelrod's model for the dissemination of culture \cite{axelrod1997dissemination}, or the well-known voter model \cite{liggett2013stochastic}. The main difference is that in our model each agent must divide its influence among all neighbours not in the same state, which is more realistic for the case of interacting cities.

At the beginning of the simulation, each city is assigned a random state (see \textbf{Fig.~\ref{fig:3}c}). Over time their states are updated under each other's influence until the whole network converges to quasi-stationarity, such that only a few states exist, each occupying a culturally cohesive core of nodes (see \textbf{Fig.~\ref{fig:3}d}). Such spatial patterns have been observed in empirical studies of human culture \cite{Blondel08}. In between the aforementioned cohesive cores there exist isolated cities that constantly flip between different states, akin to fuzzy cultural boundaries (see red boxes in \textbf{Fig.~\ref{fig:3}d}). A city's flip rate (i.e. the number of state changes per simulation iteration) can be interpreted as the magnitude of the tension attempting to change a city's culture or government. We find that the flip rate correlates strongly with betweenness $B(v)$. Moreover, the {\it modularity} (a measure of how well-defined a network's community structure is) associated with the clusters of equal state cities increases rapidly during the initial transient period and then settles at close to unity, indicating that the model detects the natural communities in the network. This is consistent with other dynamical models which have been found to reflect community structure in the same way \cite{arenas2007synchronization, boccaletti2007detecting}. Despite this model's simplicity, the fact that it displays the same threshold relationship between activity and betweenness as we have observed empirically supports the conjecture that it is the formation of fuzzy cultural boundaries between cohesive cores which mediates human conflict.
\begin{figure}[t]
    \centering
    \includegraphics[width=1.0\linewidth]{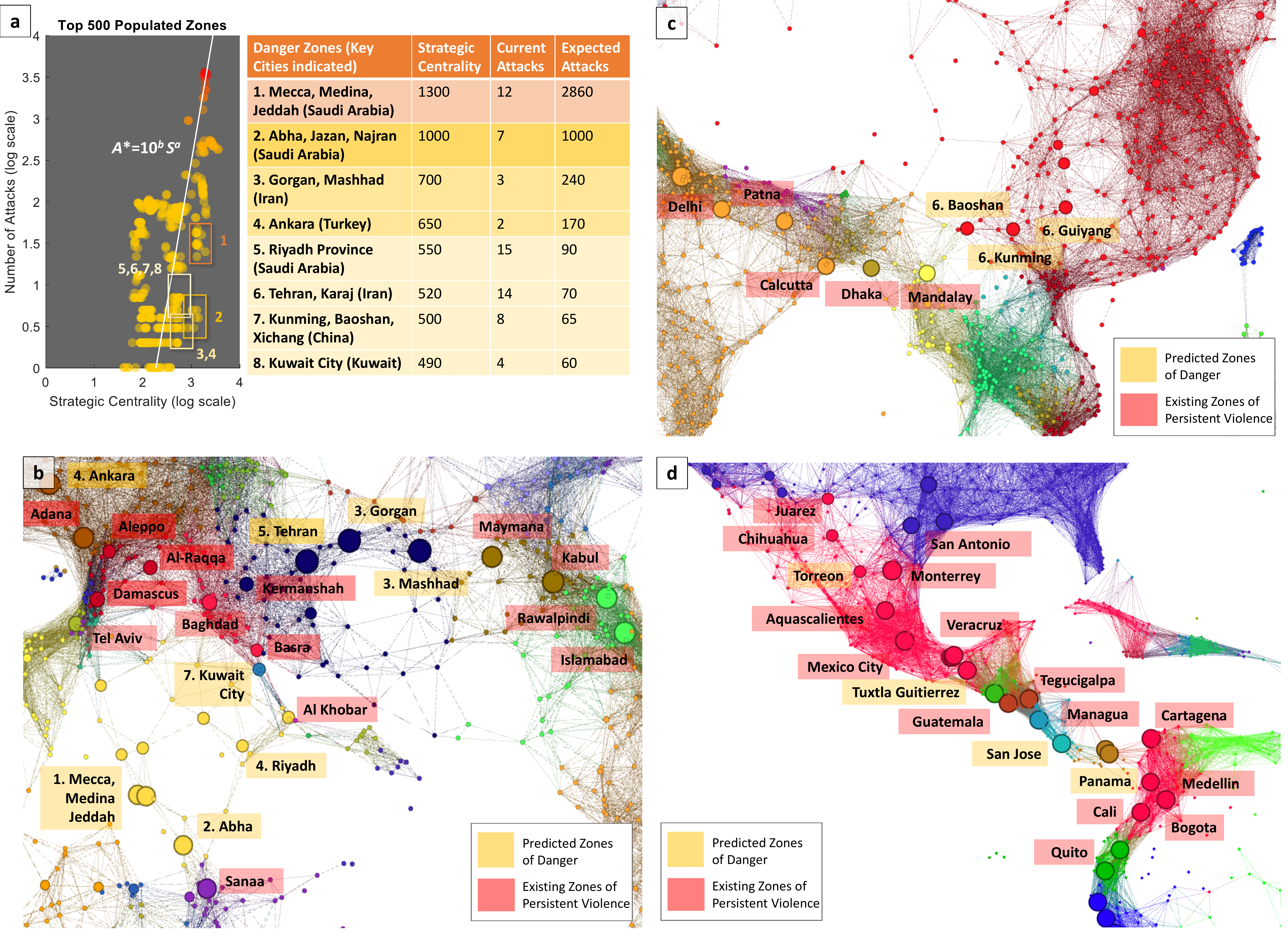}
    \caption{{\bf High population zones that are either in existing persistent violence (red labels) or are vulnerable to future violence (yellow labels).} In sub-plot a), we identify outliers zones that have a low number of attacks compared with peer zones with a similar level of strategic centrality. In the table, the current number of attacks accumulated in each zone $A(z)$ between 2002 to 2014 is presented along side the predicted number of attacks $A^{*}(z)$. Three major areas of vulnerability have been identified: b) Saudi Arabia and Iran, c) southwestern China, and d) Central Americas.}
    \label{fig:4}
\end{figure}

\subsection{Predictions}

One application of the model is to identify high risk areas that are currently relatively peaceful. \textbf{Fig.~\ref{fig:4}} highlights cities whose strategic centrality predicts a significantly higher number of terrorist attacks than currently experienced as of 2015 (\textbf{Fig.~\ref{fig:4}a}). Two major geographic areas have been identified (yellow labels): i) Saudi Arabia, Iran, Turkey (\textbf{Fig.~\ref{fig:4}b}), and ii) southwestern China (\textbf{Fig.~\ref{fig:4}c}). Of particular interest is Saudi Arabia, which is surrounded by several major existing conflict zones, i.e. Yemen, Syria and Iraq (red labels). \textbf{Fig.~\ref{fig:4}d} shows that the prediction algorithm is also accurate for the American continent, where persistent violence between international criminal organizations dominates the genre of conflict. Island networks with a violent history such as Northern Ireland and the Southern Philippines can be detected as high strategic centrality by analyzing the surrounding islands as isolated interconnected sub-networks. Using this method of prediction, 74\% of the terrorist attacks in 2016 occurred within 50km of the high betweenness cities.

\subsection{Discussion}

Tolstoy once remarked that ``individuals struggle between necessity and free-will, and the sum of individuals amounts to a general collective behaviour, which can explain war and peace.'' We have shown that the collective behaviour can be modeled as as global spatial interaction of our cities, and is statistically related to human violence. In particular, the bottlenecks of a global network of cities often display persistent conflict. We conjecture that this empirical pattern ensues from the fact that such bottlenecks often correspond to the ``fuzzy cultural boundaries'' identified in the conflict literature as risk factors. To test this hypothesis we propose a simple agent-based model of cities which can influence each other to adopt the same state as themselves (where states might represent the particular culture or government of the city). And indeed, in this model the cities in the cohesive cores of the network settle down to a single state, whereas the bottlenecks continue flipping indefinitely between states. In fact, the flip-rate exhibits the same threshold relationship to betweenness centrality as the number of attacks do in the conflict data.

Together, these results support the view that human violence is related to fuzzy cultural boundaries, which in turn reflect structural features of a network of interactions between populations. This is in keeping with previous findings: on the one hand, with the notion that fuzzy cultural boundaries cause ethno/cultural conflicts \cite{Lim07} (see Fig.~\ref{fig:2}a-i); and on the other, with the fact that conventional battles take place in central locations as in Mackinder's \textit{The Pivot of History} geopolitical theory \cite{Kaplan12, Blouet05, Venier04} (see Fig.~\ref{fig:2}j). However, while our agent-based model suggests that cultural interaction of some kind is the most likely mechanism behind the empirical findings, other potential causal explanations should \textit{not be} dismissed out of hand. 

Whilst it is true that the distribution of cities in any given region might be the result of complex historical dynamics, we should mention that the spatial bottlenecks are the result of intricate interactions among the whole connected world (i.e. the model loses predictive power when only a subset of nodes is considered). The statistical accuracy of our spatial network model does not in itself reveal \textbf{causality}, even if it serves as an elegant way of capturing the effects of such complex historical dynamics. However, unless history has caused both ancient and new cities to form in a particular spatial pattern leading to the observed close relationship between a city network and violence, it seems more likely that the simple cultural interaction model presented in the paper does indeed describe at least one dominant mechanism at work. But other mechanisms linking civilization-level network structure and ground-level human interactions should also be explored. 

As urban population grows, the emergence of new cities, transport links, and other connected infrastructures and international trade networks \cite{Cranmer_15} as interdependent multiplexed networks presents humanity with an opportunity to improve topological resilience. Dynamic effects such as mass human migration as a response to war and climate change \cite{Linkov14} also deserve attention, and we encourage others to investigate more deeply the role of spatial networks in the hope of informing policies and interacting with the politics of anthropocentric resilience. \\

\bibliographystyle{IEEEtran}
\bibliography{IEEEabrv,Ref}

\textbf{Acknowledgments}
W.G. conceived the idea and sourced the data, W.G. and X.L. designed, constructed, and analyzed the complex network. G.M.D independently verified the complex network results. S.J., G.M.D. and W.G designed the agent based model for cultural interactions. G.M.D. simulated the results for the agent based model.

The authors have no competing financial interests to report.\\
Correspondence and requests for materials should be addressed to: weisi.guo@warwick.ac.uk \\
Email address of X. Lu: luxueke9013@gmail.com \\
Email address of G. Mosquera Do\~nate: g.mosequera-donate@warwick.ac.uk\\
Email address of S. Johnson: s.johnson.2@warwick.ac.uk \\

\clearpage

\section{Material and Methods}

\subsection{Background to Conflict Modeling}

Human conflict in one guise or another has shaped our world, enthralled historians for millennia, and continues to represent an existential threat to humanity. Here we focus on modeling modern conflict, including both terrorism and conventional warfare. Whilst machine predictors are unlikely to ever replace human experience and judgment, quantitative models are useful for producing risk assessments, aid decision making, and testing hypotheses.

The quantitative models can be broadly divided into two categories. In the first category, statistical approaches can observe common ecological patterns \cite{Richardson41, Bohorquez09}, and both long-term cyclic patterns \cite{Denton68, Dewey71, Hsiang11} and short-term self-excitation mechanisms \cite{Mangion12} have been well investigated. By training machine-learning algorithms on historical patterns, accurate predictions (70-80\%) on new emerging conflicts can be made. Such algorithms are used by commercial (e.g. Parus Analytics) and government (e.g. US Department of Defence - Integrated Crisis Early Warning System) organizations. However, the underlying algorithms are sensitive to data availability, lack detailed quantitative mechanisms for sociopolitical hypotheses testing, and as such find it difficult to provide solutions. In the second category, Agent-Based Models (ABM) has the benefit of being able to simulating the impact of specific sociopolitical causal mechanisms, such as culture clashes \cite{Lim07}, climate change \cite{Zhang07}, or technology transfer \cite{Turchin_13}. However, ABMs require detailed configurations for each geographic region, and are often too complex for creating a universal understanding of conflict. It is also worth mentioning spatial interaction models that sit in between the aforementioned approaches \cite{Wilson08}, and have proven predictive power in a variety of topics including threat projection between conflicting parties \cite{Baudains15}. Despite the accuracy of these modelling approaches, there lacks a bridge between them. 

\subsection{Data Set and Network Construction}

The paper leverages on the following open data sets:
\begin{enumerate}
  \item \textbf{City data}: 7322 cities with their latitude, longitude, and population data from National Geospatial Intelligence Agency \footnote{The dataset has been improved with population data and can be found at \cite{Cities_Data}}. The geospatial data includes cities that vary in population from mega-cities (several millions) to small towns. The data represents $\approx$25\% of the world's total population and it includes over 2800 cities with a population over 100,000, yielding a sufficiently high city resolution. For the purpose of this paper, we shall call all settlements \textit{cities}. Each city is also tagged with its country and province affiliation.
  \item \textbf{Conflict data}: two data sets are used, (1) terrorism and insurgency violence, and (2) conventional warfare. For terrorism and insurgency violence: the Global Terrorism Database (GTD) \cite{GTD} database is used, with over 30,000 conflict incidents between 2002 to 2014. The GTD contains the number of attacks and death-toll, ranging from small-scale assassinations (1 death) to large-scale massacres (1000s dead). Most of the death-toll data is time stamped and geo-tagged (longitude and latitude). The GTD data is further processed to: (i) fill in any missing longitude and latitude information, (ii) remove attacks with unknown location information (less than 1\%), and (iii) cluster the number of attacks and death-toll data to the nearest city in the geospatial data set (mean distance 27km). For conventional warfare: the PRIO - Uppsala Conflict Data Program (UCDP) database is used \cite{PRIO}, with over 150 conventional battles. Most of the battles are tagged with the location. Further manual geo-tagging of GPS location was performed by the authors.
  \item Geopolitical and socioeconomic data: the 2014 GDP per capita and income inequality data from the World Bank and the International Monetary Fund, and the democracy index developed by the Economist Intelligence Unit (EIU).
\end{enumerate}

The aim of this paper is to develop an approximate global interaction model based on land connections between cities. It is well established that the interaction weight between cities (i.e., transport flow volume) can be approximately measured by the gravity law. The general equation states that the symmetrical flow between two cities $i$ and $j$ is proportional to the population of the two cities $P$ and inversely proportional to the square of the Euclidean distance $d$, such that: $F_{ij} = \frac{P_i P_j}{{d_{ij}}^{2}}$ \cite{Road_Traffic_07, Highway08} (see Supplementary Information (SI) for further justification). As this would potentially create a land route between any two cities worldwide, we add a hard-disk radius of $500$km as the constraint for the longest land route between any two settlements. This way, a multi-hop interaction network is created. Two further constraints are put in place: a) eliminate sea travel ($>50$km), and b) in order to only consider major links and cities, we only consider cities with a population over 10,000 people, which are less sensitive to population changes (83\% of all cities in the data set). The latter 2 constraints are not strictly necessary to achieve similar statistical results. As a result, a global multi-hop interaction network is constructed (see Fig.~\ref{fig:1}b). 

\subsection{Top-Down Complex Network Analysis}

As conflicts often happen over an area involving multiple cities, we are interested in the properties of a zone $z$. We define a zone as an area that has many cities, i.e., $v \in V_{z}$. In the main paper, the results are presented for circular zones of 500km in radius, each representing 0.5\% of the modeled global land surface area. Overlap of the zones is permitted and each zone is centred on a particular city. The paper is firstly interested in two primary measures in network science: degree and betweenness. A city's \textit{degree} is defined as the number of links with neighbouring cities, $D(v) = \text{deg}(v)$. The degree is unweighted to highlight the importance of the number of links (neighbours), as opposed to the importance of the neighbours or links. The total degree property of a zone $z$ is defined as the sum of all the degree values of each city inside the zone (number of links inside a zone), i.e., $D(z) = \sum_{v \in V_{z}} D(v)$. Interaction routes often travel the shortest multi-hop path between a number of cities. Betweenness measures the number of shortest paths through a node (i.e., city). The shortest path of travelling between a city $m$ to any other city $n$ is defined as the path with the least number of hops. The unnormalised betweenness of a city $B_{v}$ is defined as the number of shortest paths that pass through this city:
\begin{equation}
B(v) = \sum_{m \neq n \neq v} \sigma_{mn}(v),
\label{eq:betweenness}
\end{equation} where $\sigma_{m,n}(v)$ is the shortest path between $m$ and $n$ that goes through city $i$. The total betweenness property of a zone $z$ is defined as the sum of all the betweenness values of each city inside the zone (number of shortest paths that pass through the zone), i.e., $B(z) = \sum_{v \in V_{z}} B(v)$. Unlike the unweighted degree, we are interested in the weight contribution of each link. This is because multi-hop interaction routes are likely to consider a balance between: i) the shortest path, and ii) one that also passes through major cities for increased profit (i.e., attracted to population $P$). For the \textbf{weighted network}, one can define the weight of each link between 2 arbitrary cities $i$ and $j$ as inversely proportional to the expected flow value \cite{Opsahl10}:
\begin{equation}
w_{ij} = F_{ij}^{-\theta},
\label{eq:weight}
\end{equation} where $F_{ij}$ is the flow value and $\theta = [0,1]$ can offer full weighting ($\theta=1$) or no weighting ($\theta=0$). For example, if the flow weight of a link is high, then the weight of the link (for shortest path calculation) is negligibly small and there is no minimum or maximum link weight. In this paper's main analysis and results section, we have selected to present the results for a balanced weighted betweenness centrality measure ($\theta = 0.5$), such that a balance exists between the weight of a link and the number of links. Nonetheless, we present the results for $\theta = 0$ (no weighting) and $\theta = 1$ (full weighting) in the SI to demonstrate the robustness of the methodology.
\begin{figure}[t]
    \centering
    \includegraphics[width=1\linewidth]{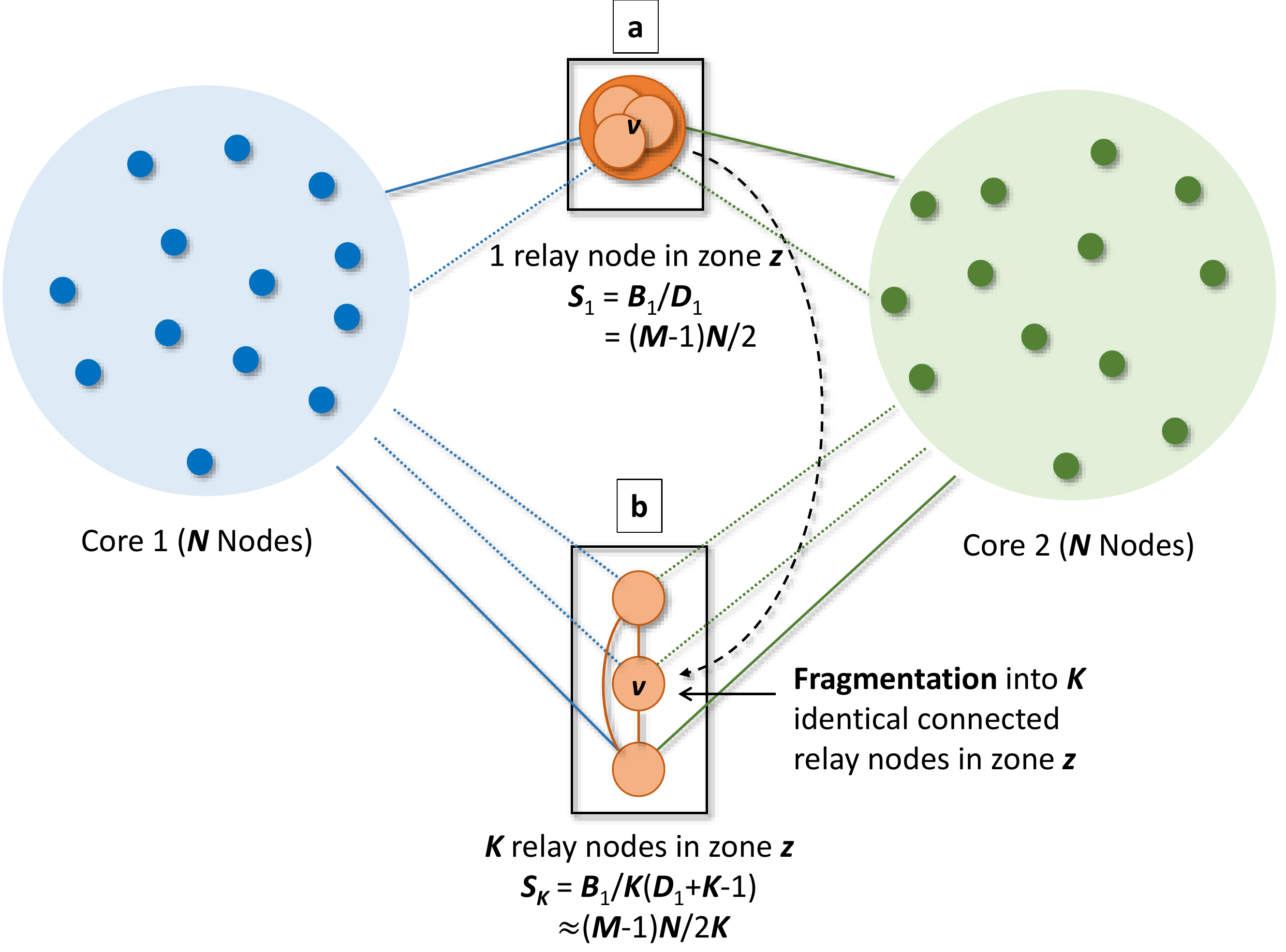}
    \caption{{\bf A graph showing $K$ relay nodes connecting $M$ cores, each with $N$ nodes.}  Scenario (a) shows a single relay node and Scenario (b) shows it fragmented into $K$ relay nodes. The solid lines indicate the shortest path and the dashed lines indicate other paths (not shortest). Each relay node in a zone of $K$ relay nodes has a degree $D_{K}(v) = MN+K-1$, a betweenness $B_{K}(v) = {M\choose 2}N^{2}/K$, and a strategic centrality $S_{K} \approx MN/2K$. In terms of cities, fragmentation of one city into many cities will reduce the betweenness and strategic centrality of any individual city and hence reduce its attraction to militants (i.e., fewer interaction routes). Fragmentation will also increase the number of neighbours and its degree, which reduces its vulnerability.}
    \label{fig:M1}
\end{figure}

The strategic centrality for a zone $z$ as $S(z) = \frac{B(z)}{D(z)}$, which normalises the betweenness of a city by the number of links. It is a measure of the number of shortest paths per link connected to a city (similar to link betweenness that is used to measure the importance of road networks \cite{Lammera06}). A city with a high betweenness will have global interaction importance, while a low degree (more isolated) will increase its vulnerability to conflict. Let us consider a simple problem: what happens to the degree and betweenness properties of the relay nodes, when the number of relay nodes $K$ changes. Let us at first consider a single relay node $v$ in a zone $z$ (scenario (a) in Fig.~\ref{fig:M1}). The degree of the node or the zone is $D_{1}(v) = D_{1}(z) = 2N$ if the relay node is connected to all $N$ nodes in the cores. The betweenness of the node or the zone is $B_{1}(v) = B_{1}(z) = N^{2}$, if every shortest path passes through the relay node. Consider now the case the single node fragmenting into $K$ relay nodes - see scenario (b), the degree value of each node becomes $D_{K}(v) = D_{1} + (K-1)$, of which there are $K$ in the zone (i.e., $D(z) = KD(v)$). As for the betweenness, one needs to consider the average value across all relay nodes in the zone, as each single relay node will have a different number of shortest paths that pass through it. In total, the same $B_{K}(z) = B_{1}$ shortest paths pass through $K$ relay nodes in zone $z$. Hence, the average betweenness in the relay nodes is $\mathbb{E}[B_{K}(v)] = B_{K}(z) = \frac{B_{1}}{K}$. Note, no shortest paths between relay nodes exist, as all relay nodes are directly connected to each other. Strategic centrality is the same for each node and for the whole zone: $S_{K}(v) = S_{K}(z) = \frac{\mathbb{E}[B_{K}]}{D_{K}} = \frac{B_{1}}{K(D_{1}+K-1)}$. More generally, if there are $M$ cores, each with $N$ nodes, the degree of a relay node is:
\begin{equation} \begin{split}
D_{K}(v) = MN + (K-1),
\label{eq:degree_example}
\end{split}\end{equation} and the betweenness of a relay node is:
\begin{equation} \begin{split}
B_{K}(v) = {M\choose 2}\frac{N^{2}}{K}.
\label{eq:between_example}
\end{split}\end{equation} Therefore, the strategic centrality of the $K$ interconnecting relay nodes is
\begin{equation} \begin{split}
S_{K} = {M\choose 2}\frac{N^{2}}{K(MN+K-1)} \approx (M-1)\frac{N}{2K},
\label{eq:strategic_example}
\end{split}\end{equation} for a larger number of core nodes compared to the relay nodes ($N \gg K$). Therefore, increasing the number of relay nodes $K$ will reduce the strategic importance of any one relay node, and increasing the number of core nodes $(M-1)N$ will increase the relay nodes' strategic importance. The strategic centrality analysis is equally applicable on the individual node $v$ level as well as the zone $z$ level.

\subsection{Bottom-Up Agent-Based Simulation}

We model individual people behaving as a collective group when they share a common culture \cite{Ellemers12}. We consider cities in a network with adjacency matrix $\{A_{ij}\}$. We also define a number of distinctive cultural states, and at any given time $t$, each city can be in any of $S$ states: $s_{i}(t) = \{1,2,3...Q\}$. We now model the willingness for the people in cities to influence other cities that it is connected to. The agent-based-model (ABM) simulates the cultural influence interaction between cities, with each city exerting its cultural influence to neighbouring cities until they agree on a common culture. This approach has been used before in a variety of ways, including fostering cooperation \cite{Johnson15}. The steps in methodology and analysis for the ABM is listed below.

\textbf{Step 1 - No. of Dissimilar Neighbours:} A way to define the number of connected cities with different cultures is: $n_{i}(t) = \sum_{j} A_{ij}[1-\delta_{s_{i}(t),s_{j}(t)}]$, where $\delta_{s_{i}(t),s_{j}(t)}$ is the Kronecker delta function (i.e., $d_{s_{i}(t),s_{j}(t)}=1$ only if the cultures are identical $s_{i}(t)=s_{j}(t)$).

\textbf{Step 2 - Projection of Cultural Influence:} We assume that each city has a finite capacity to exert influence to neighbouring cities with a different culture. This capacity $C_{j}$ could be related to total population, military strength, wealth...etc. In this paper, we assume it scales linearly with population. As such, the amount of influence $f_{j}$ is shared amounts the $n_{j}$ connected cities with different cultures such that:
\begin{equation}
f_{j} = \frac{C_{j}}{n_{j}(t)}.
\label{eq:influence}
\end{equation} As such, city $i$ is influenced by culture from neighbouring cities $j$ that have a culture $s_{j}=q$. The weight of this cultural influence is:
\begin{equation}
h_{i}^{q}(t) = \sum_{j} A_{ij}\delta_{q,s_{j}(t)}f_{j}.
\label{eq:influence2}
\end{equation} 

\textbf{Step 3 - Adoption of a Culture:} Due to the influence, city $i$ will adopt a culture $q$ at time $t+1$ with probability:
\begin{equation}
P(s_{i}\rightarrow c) \propto [h_{i}^{q}(t)]^{r},
\label{eq:influence3}
\end{equation} where $r$ can be seen as a \textit{rationality} factor. Low rationality ($r=0$) leads to random cultural adoption and high rationality leads to switching to the cultural state $q$ with the strongest influence $h_{i}^{q}(t)$.

\textbf{Step 4 - Monte Carlo Simulation:} This simulation starts with populating each city uniformly and randomly, and then connecting them according to the hard disk radius. The edges of the simulation are spatially wrapped to remove boundary effects. At the start, each city has a random distinctive culture and converges after some time to form culturally cohesive city cores (see Fig.~\ref{fig:1}a-iii). Yet, there are many cities that do not converge on a singular culture and flips between multiple cultures, or can otherwise be seen to be part of a fuzzy culture boundary (see Fig.~\ref{fig:1}a-iii). The process is repeated until results converge.

\textbf{Step 5 - Analysis:} As reflected in the results (see Fig.~\ref{fig:3}b), the cities that converge on a high cultural flip rate also have high network betweenness $B(v)$ and the cities that converge to a cohesive culture with its neighbours also have a high degree $D(v)$.

\end{document}